# Prioritizing the Components of Vulnerability in a Genetic Algorithms Minimization of Flood Risk

Vena Pearl Boñgolan
Karessa Alexandra O. Baritua
Marie Junne Santos

Scientific Computing Laboratory
Department of Computer Science, College of Engineering,
University of the Philippines Diliman

## ABSTRACT
We compare two prioritization schemes for the components of flooding vulnerability: urbanized area ration, literacy rate, mortality rate, poverty, radio/tv penetration, non-structural measures and structural measure. We prioritize the components, giving each a weight. We then express the vulnerability function as a weighted sum of its components. This weighted sum serves as the fitness function in a genetic algorithm, which comes up with the optimal design for a flood-resistant city.

## Categories and Subject Descriptors
I.2.8 [**Problem Solving, Control Methods, and Search**] genetic algorithms, multi-objective optimization.

## General Terms
Algorithms

## INTRODUCTION
There are many factors as to why the Philippines experiences continual flooding in various areas during the rainy season. The purpose of conducting a Risk Assessment Analysis is to know the threat of a damage, liability, loss, or other negative occurrence caused by internal or external vulnerabilities. It is also used to identify the flood risk in certain areas using risk factors such as the area's urbanized area ratio, literacy rate, mortality rate, percent of population under poverty, radio/TV penetration and the state of structural and non-structural measures

We would like to study these factors throughly so we could identify the characteristics of a city with minimal flooding risk. This could serve as a guide for community developers in their city plans. Also, various government agencies can refer to our paper for their action plans and reinforcements towards the improvement of different communities. There can be an overall increase in community preparedness for disasters, and better infrastructure can be built.

## BACKGROUND

Genetic Algorithms (GAs) have gained immense popularity in real-world engineering search and optimization problems. In the field of flood management, GAs have been used in the design of flood control structures (Wallace & Louis, 2003), model calibration (Lan, 2001), flood plain management (Karamouz, Abesi, Moridi, & Ahmadi, 2009), and flood forecasting (Mukerji, Chatterjee, & Raghuwanshi, 2009). This study aims to expand this list to include flood disaster management. GAs are computerized search and optimization methods which mimic the principles of natural evolution. Based on Darwin's survival-to-the-fittest design solutions, GA's intelligent search procedure finds the best and fittest design solutions, which are difficult to find using other techniques. Because of this problem's high dimensionality, similar to multi-objective planning problem tackled by Balling et al. in [1], direct methods are intractable if not impossible to use, genetic algorithms naturally present themselves.

All disaster management plans begin with the identification of risk (UN/ISDR, 2004), see [2]. This is true whether we are concerned with earthquakes, storms, floods or other natural hazards. Risk is an indicator of how prone a specific area is for a natural hazard to turn into a disaster and is a function of three factors which is defined by this equation:

Risk = Hazard x Vulnerability x Exposure

When natural hazards are concerned, little can be done on the hazard probability of a certain area for a specific disaster; this is due to the fact that it is predominantly defined by geographic location. Manipulating exposure, on the other hand, would mean manipulating the number of people or control if not impede any kind of growth in the area.This is why most disaster management plans focus on modifying the vulnerability index to minimize risk. That is how we intend to approach this study.

## METHODOLOGY

Like the previous research [3], the study area will be divided into several computational units that can be defined by the political boundaries, e.g., barangays (communities or neighborhoods within a city). A set of plans are to be randomly generated called the "starting generation." These will represent various designs of the city and will be used to produce the succeeding generations. In the final generation, the chromosomes of the optimal city will be interpreted.

A 6 by 6 grid will be used to represent the city and its 36 areas/barangays.

| ½ | ½ | 1 | 1 | 2 | 2 |
|---|---|---|---|---|---|
| ½ | 1 | 1 | 2 | 2 | 2 |
| 1 | 1 | 2 | 2 | 2 | 1 |
| 1 | 2 | 2 | 2 | 1 | 1 |
| 2 | 2 | 2 | 1 | 1 | ½ |
| 2 | 2 | 1 | 1 | ½ | ½ |

**Figure 1. 6x6 grid representing a city**

The numbers inside the cells represent physical properties of the area which might make it more, or less, vulnerable to flooding, and will be used to multiply the area's vulnerability. Here, we may imagine a river running diagonally down the grid, from upper right to lower left. Hence, the barangays on the right diagonal lie on a flood plain, and will have a factor of two (the danger zone). We assume the terrain rises a bit above the flood plain on either side of the diagonal, and have a value of 1. Those barangays farthest from the floodplain, and possibly on higher ground, we give a factor of ½.

The fitness function for the GA will be composed of the following components:

Urbanized area ratio – Non-urbanized areas are devoid of infrastructure and residents. The less people and infrastructure affected, the lower the vulnerability to flooding.

  11 = Highly urbanized
  10 = Moderately urbanized
  01 = A little urbanized
  00 = Not urbanized

Literacy Rate – Here we take the definition of literacy as being able to read warning signs from the government, and being able to understand instructions from emergency responders. This has implications on over-all community preparedness.

A high illiteracy rate increases the vulnerability to flooding.

  11 = more than 75% are illiterate
  10 = 50 – 75% are illiterate
  01 = 25 – 50% are illiterate
  00 = 0 – 25% are illiterate

Mortality Rate – We may take this as an indicator of the over-all health of the community. The higher the existing mortality rate, the more likely people would be adversely affecter if a disaster were to occur.

  11 = High mortality rate
  10 = Average mortality rate
  01 = Below average mortality rate
  00 = Low mortality rate

Population under poverty – Community members living in poverty would have a hard time coping with disasters, not only in the immediate aftermath, but in the recovery process as well. While floods tend to affect wide areas, and all socio-economic classes in an area, we expect those community members with more resources to have a shorter recovery time after disaster strikes, not to mention more options which might include moving to other areas. The more members there are living below the poverthe ty line, the more vulnerable an area is.

  11 = more than 75% in class D
  10 = 50 – 75% in class D
  01 = 25 – 50% in class D
  00 = 0 – 25% in class D

TV / Radio penetration rate – The lower the TV/Radio penetration rate is, the less likely the people would receive the news and warnings that are broadcast. We also see TV/Radio as important components in disaster preparedness. This would mean that areas with low mass communications penetration would not be able to prepare themselves for incoming disasters, thus, increasing the vulnerability.

  11 = less than 25% penetration rate
  10 = 25 – 50% penetration rate
  01 = 50 – 75% penetration rate
  00 = 75 – 100% penetration rate

State of non-structural measures – These are laws legislated, and, more importantly, compliance to laws that were meant to minimize flood risks, e.g., respecting te three-meter easement on creeks and waterways, cleaning drainages, litter prevention. The greater the compliance, the less vulnerable an area will be.

  11 = no non-structural measure
  10 = existing with poor implementation/compliance
  01 = existing with average implementation/compliance
  00 = existing with good implementation/compliance

State of structural measures – These are the infrastructure built to prevent and/or control floods. (i.e. drainage, flood gates, pumping stations, etc.). The better-maintained infrastructure is, the easier for floodwaters to receed from an area, if ever flooding were still to occur. This would, lessen the area's vulnerability.

11 = no structural measure
10 = existing structural measure in poor condition
01 = existing structural measure in average condition
00 = existing structural measure in good condition

| Hazard Aspect | Always Very Important | Usually Important | Sometimes Important | Rarely of Importance | Not Worth Considering |
|---|---|---|---|---|---|
| Likelihood of Occurrence | | | | | ✓ |
| Capacity to cause physical damages | ✓ | | | | |
| Size of Affected Area | | | ✓ | | |
| Speed of Onset (amount of warning time) | | | | ✓ | |
| Percent of population affected | ✓ | | | | |
| Potential for causing casualties | ✓ | | | | |
| Potential for negative economic effects | | ✓ | | | |
| Duration of threat from hazard | | | | ✓ | |
| Seasonal risk pattern | | | | | ✓ |
| Environmental impact | ✓ | | | | |
| Predictability of hazard | | | | ✓ | |
| Ability of hazard impacts to be mitigated | | | | ✓ | |
| Availability of warning systems | | | ✓ | | |
| Public awareness of hazard | | ✓ | | | |
| Corollary effects (ability to cause other hazards) | | | ✓ | | |
| (Other considerations may be added to this list) | | | | | |

**Fig. 2. Vulnerability Determination Table**

Figure 2 comes from the analysis of the state of Michigan, USA, see [4]. They give a list of aspects that are to be considered for a hazard (here flooding), and we classify the aspects according to their importance. We then assigned percentages to these aspects. The ones marked "always very important" were each given 20%, and the ones marked "usually important" 10% each. All the other columns were ignored in this study. There are no strict rules on how points are to be assigned, which perhaps makes this analysis subjective.

The first aspect is the capacity of the flood to cause physical damages. It is important to study the severity of the disaster that will hit our country anytime, so that proper avoidance and prevention schemes shall be implemented in the future. The second aspect is the percentage of the population being affected by the flood. The third aspect is the potential for causing casualties. People's lives can be at stake because of the many consequences of floods everywhere. Knowing the potential deaths and injuries caused by the floods can help certain government agencies to address them properly. Search and rescue operations will be managed more effectively in the future. The fourth aspect is the potential for negative effects. This is very considerable, since we have a large number of businesses all over the country. Because floods cause damage in physical properties, a decline in the aspect of economy is also possible. There could be challenges in rebuilding these properties. Shortage in food is also a probability. This can be the root of price hikes. The fifth aspect is the environmental impact of the flood. Floods can cause a big havoc to the environment. Shortage of food crops can be caused due to loss of entire harvest. Trees can die of suffocation. The overflow of water can submerge a large amount of land, ruining the plant life entirely, at the worst case scenario. This is something we have to avoid, so that our environment can still survive. The last aspect that we considered is the public awareness hazard. Literacy rate is a very vital factor in determining public awareness. Individuals may not be properly warned due to their incapacity to read or understand cautionary signs in their specific zones. Some people may be unaware of the disasters about to hit them because of a lack of radio and TV penetration in their areas as well.

| | Capacity to cause physical damages (20%) | Percentage of the population affected (20%) | Potential for casualties (20%) | Environmental Impact (20%) | Negative economic impacts (10%) | Public awareness of hazard (10%) | TOTAL |
|---|---|---|---|---|---|---|---|
| Urbanization | 9<br>-The more urbanized a community is, the more likely for the physical damages to occur. | 7<br>-More people will be affected if the area is more urbanized. | 5<br>-If the area is more populized, the potential for casualties would be more likely to increase but not everyone will get injured or die. | 4<br>-No matter how urbanized a community, the effect of flood is out of our hands. | 4<br>-More urbanized area will cause less economic loss in the community. | 5<br>- With an urbanized area, the dissimenation of information would be more easier. | (9.0 x 0.20) + (7.0 x 0.20) + (5.0 x 0.20) + (4.0 x 0.20) + (4.0 x 0.10) + (5.0 x 0.10)= **6.3** |
| Poverty | 4<br>-The less fortunate the community, their would be less physical damage that will occur. | 6<br>-If a community is less fortunate it will be more likely that the percentage of the population affected will be high. | 5<br>-Being less fortunate might increase potential for casualty. | 4<br>-No matter how less fortunate a community is the effect of flood in the environment is uncontrolable. | 3<br>-Since the community is less fortunate, there will be less economic impact that will occur | 5<br>-If they are not equiped with materials that will help them be informed, it will be harder for them to know what is happening. | (5.0 x 0.20) + (6.0 x 0.20) + (6.0 x 0.20) + (4.0 x 0.20) + (3.0 x 0.10) + (5.0 x 0.10)= **5.0** |
| Literacy Rate | 5<br>-The better the people can understand/ read warning signs, the more prepared they will be in responding emergency situations as such. | 4<br>-Lesser percentage of population will be affected since the people are more ready. Although, a number can still be affected for the reason that the effects of the flood are out of their control. | 3<br>-In an area, a number of people may be affected, but not all are likely to be injured or die, in the worst-case scenario | 6<br>-People may be able to read and understand warning signs, but the destruction of the environment is out of our hands. | 4<br>-Knowing the possible effects that the disaster may bring beforehand, certain preventive measures for economic loss can be planned and managed effectively. | 9<br>-With the capability of the community to read and understand warning signs, they can be well aware of the possible hazards that the flooding can bring. | (5.0 x 0.20) + (4.0 x 0.20) + (3.0 x 0.20) + (6.0 x 0.20) + (4.0 x 0.10) + (9.0 x 0.10) = **4.9** |
| Mortality Rate | 4<br>-The number if deaths in a community will not affect much the physical damages that might occur. | 9<br>-More people will be affected if the mortality rate is high. | 7<br>-If there is high mortality rate in a community, there is a posibilty that casualty will be high. | 5<br>-The environmental effect is sometimes unpredictable. | 6<br>-High mortality rate will be a big loss causing high negative economic impact. | 5<br>-Being well informed regarding flood and its effect will be of great help to the community. | (4.0 x 0.20) + (9.0 x 0.20) + (7.0 x 0.20) + (5.0 x 0.20) + (6.0 x 0.10) + (5.0 x 0.10) = **6.1** |
| TV/Radio Penetration | 5<br>-The greater the TV and radio penetration, the less likely it would cause physical damages, since certain preventive measures can be implemented. | 4<br>-There will be a lesser percentage in the population being affected, since people are more prepared on how to act in these emergency situations | 3<br>-Fewer people are less likely to be injured or die. | 5<br>-The environmental damage that can be brought by flooding is sometimes out of our hands. No matter how aware we can be, the environment can still be greatly affected. | 3<br>-Through the warnings given via the radio or TV, people may act immediately on their business activities, thereby preventing the decline of economy. | 10<br>-The public awareness of people prior to the flooding is contributed by the warnings heavily emphasized in TV, radio and other means of communication. | (5.0 x 0.20) + (4.0 x 0.20) + (3.0 x 0.20) + (5.0 x 0.20) + (3.0 x 0.10) + (10.0 x 0.10) = **4.7** |
| Non-structural Measures | 5<br>-If people would be doing their responsibilities in terms of complying with the laws to prevent flooding, there would be less chances that it would cause physical damages. | 4<br>-A minimization of flood risks is possible since people abide by the laws legislated such as cleaning drainages and litter prevention. | 3<br>-With a smaller percentage of population being affected, there would be an even lesser potential for casualties. | 3<br>-People may follow the laws reinforced to them, but natural disasters are out of our control, and the effect it can give to the environment are very much unpredictable. | 2<br>-Compliance to the laws by people can have a lesser negative impact on the economy than that of the environment, since we can still control some businesses we have through flood prevention schemes. | 6<br>-With the whole community following the flood prevention laws imposed to them, and having the discipline to do this, they can be well aware of the possible effects of their actions. | (5.0 x 0.20) + (4.0 x 0.20) + (3.0 x 0.20) + (3.0 x 0.20) + (2.0 x 0.10) + (6.0 x 0.10) = **3.8** |
| Structural Measures | 2<br>-With better infrastructure built to minimize flooding, there could be less damage to physical properties, and it would lessen the vulnerability of an area. | 3<br>-Since better buildings, homes, infrastructure are built, fewer people can be affected. | 2<br>-With a smaller percentage of population being affected, there would be an even lesser potential for casualties. | 4<br>-Having flood gates, pumping stations and the like to minimize flood could also save the environment. However, some destructions can still be out of our hands. | 3<br>-The development of structural measures can contribute to a better economy. Lesser recovery systems shall be implemented because there will also be lesser areas to be affected; thereby minimizing the cost for vulnerabilities. | 5<br>-The enhancement of our infrastructure can very well inform the public on what to do to maintain these structures for the future. This can let them be aware that such measures are for their benefit prior to the flooding effects that they may experience. | (2.0 x 0.20) + (3.0 x 0.20) + (2.0 x 0.20) + (4.0 x 0.20) + (3.0 x 0.10) + (5.0 x 0.10) = **3.0** |

**Fig. 3. Table** Analysis

After narrowing the questions to the six we consider most significant, and giving them points (totalling 100), we proceeded to fill the second table (see [4]) by listing our set of seven components, namely urbanization, poverty, literacy rate, mortality rate, radio/ TV penetration, non-structural measures and structural measures as rows, and the six questions as the columns.. We then interrogated each component of vulnerability, giving a rating from 1 to 10. Our reasons for doing so are shown in the table above. From this process, we have now come up with our weights as follows:

Urbanization: 6.3/33.8
Literacy: 4.9/33.8
Mortality: 6.1/33.8
Poverty: 5.0/33.8
Radio/ TV Penetration: 4.7/33.8
Non- Structural Measures: 3.8/33.8
Structural Measures: 3.0/33.8

A previous analysis done following Einarsson and Rausand [5] came up with the following weights:

Urbanization: 3/26
Literacy: 3/26
Mortality: 3/26
Poverty: 5/26
Radio/ TV Penetration: 5/26
Non- Structural Measures: 5/26
Structural Measures: 2/26

These weights were used in the same equations to allow comparison between the results.

As with previous research, [3], we define the cost function as a 'penalty': we take the three's complements of the chromosome, then assume exponential growth for 'expensive' activities like urbanization, poverty alleviation and building structural measures.
All other chromosomes are assumed to have linear penalties, except the mortality variable, which we assume to be quadratic. As mentioned earlier, the chromosomes are independent, corresponding to independent solutions and costs, but we allow interactions between poverty and mortality, and literacy and radio/TV penetration.
We simulated these equations separately – the first and second vulnerability function with the same cost function.

## RESULTS AND DISCUSSIONS
We input the chromosomes into MatLab's genetic algorithms toolboxes. The results we got from running GA in the toolbox produced the following ideal characteristics for the 36 barangays, broken-down by chromosomes as:

1. Urbanization: The area in the diagonal shows that half of them are urbanized and half are not.
2. Poverty: The algorithm placed more less fortunate area on the diagonal.
3. Literacy: More illiterate area are placed on the diagonal.
4. Mortality: The algorithm placed area with low mortality rate on the diagonal.
5. Radio/ TV Penetration: More than half of the area in the diagonal have high penetration.
6. Non-Structural Measures: The diagonal shows that half of the area have good laws for flood while some doesn't.
7. Structural Measures: Most of the area in the diagonal have no structural measure, caused by high means for having good structural measure.

**Urbanization**
11, most urban

| 00 | 10 | 01 | 10 | 00 | 00 |
|----|----|----|----|----|----|
| 00 | 11 | 11 | 10 | 10 | 01 |
| 01 | 01 | 10 | 11 | 10 | 11 |
| 01 | 10 | 00 | 10 | 10 | 00 |
| 00 | 01 | 00 | 00 | 10 | 00 |
| 10 | 00 | 01 | 10 | 11 | 10 |

**Poverty**
11, more people under poverty

| 01 | 00 | 01 | 11 | 11 | 10 |
|----|----|----|----|----|----|
| 10 | 00 | 01 | 01 | 11 | 11 |
| 10 | 00 | 00 | 10 | 11 | 01 |
| 10 | 01 | 00 | 01 | 00 | 11 |
| 00 | 10 | 00 | 10 | 10 | 01 |
| 10 | 10 | 10 | 10 | 11 | 00 |

**Literacy**
11, most illiterate

| 00 | 00 | 00 | 11 | 11 | 00 |
|----|----|----|----|----|----|
| 00 | 01 | 10 | 10 | 11 | 01 |
| 11 | 00 | 10 | 10 | 10 | 10 |
| 01 | 00 | 00 | 00 | 10 | 01 |
| 10 | 00 | 10 | 01 | 11 | 01 |
| 00 | 00 | 11 | 10 | 10 | 00 |

### Mortality
11, highest mortality

| | | | | | |
|---|---|---|---|---|---|
| 01 | 11 | 00 | 01 | 00 | 00 |
| 10 | 01 | 00 | 11 | 10 | 00 |
| 11 | 11 | 01 | 11 | 00 | 10 |
| 00 | 01 | 01 | 10 | 11 | 10 |
| 00 | 01 | 00 | 11 | 00 | 11 |
| 00 | 01 | 01 | 00 | 01 | 11 |

### Radio/ TV Penetration
11, least penetration

| | | | | | |
|---|---|---|---|---|---|
| 01 | 11 | 00 | 11 | 10 | 00 |
| 11 | 10 | 01 | 01 | 01 | 01 |
| 01 | 10 | 11 | 11 | 10 | 01 |
| 01 | 11 | 00 | 10 | 01 | 11 |
| 00 | 00 | 01 | 10 | 01 | 00 |
| 10 | 10 | 01 | 01 | 01 | 10 |

### Non-Structural Measures
11, no laws/ poor compliance

| | | | | | |
|---|---|---|---|---|---|
| 00 | 01 | 01 | 01 | 00 | 10 |
| 01 | 11 | 10 | 00 | 01 | 01 |
| 10 | 01 | 00 | 11 | 01 | 01 |
| 11 | 00 | 01 | 00 | 11 | 01 |
| 01 | 11 | 10 | 00 | 10 | 00 |
| 00 | 00 | 10 | 10 | 10 | 11 |

### Structural Measures
11, no structural measure

| | | | | | |
|---|---|---|---|---|---|
| 10 | 01 | 01 | 01 | 01 | 11 |
| 11 | 10 | 10 | 10 | 11 | 11 |
| 01 | 10 | 10 | 11 | 00 | 10 |
| 11 | 01 | 10 | 01 | 00 | 01 |
| 10 | 10 | 00 | 00 | 10 | 10 |

| 01 | 11 | 01 | 11 | 10 | 11 |
|---|---|---|---|---|---|

As for the past study's weights used these are the resuts:

1. Urbanization: Half of the area in the diagonal are urbanized and half are not.
2. Poverty: The program produced a model with more capable/rich people in the diagonal.
3. Literacy: The diagram shows that more literate barangays are placed in the diagonal.
4. Mortality: More barangays with high mortality rate were placed in the diagonal.
5. Radio/ TV Penetration: More than half of the area in the diagonal have least penetration.
6. Non-Structural Measures: More than half of the diagonal have good law for flood.
7. Structural Measures: Mostly of the barangays placed in the diagonal have no structural measure.

### Urbanization
11, most urban

| | | | | | |
|---|---|---|---|---|---|
| 11 | 11 | 00 | 10 | 01 | 11 |
| 00 | 10 | 01 | 10 | 01 | 01 |
| 10 | 00 | 00 | 01 | 01 | 11 |
| 11 | 11 | 00 | 01 | 00 | 11 |
| 01 | 10 | 00 | 11 | 10 | 01 |
| 11 | 01 | 11 | 01 | 01 | 11 |

### Poverty
11, more people under poverty

| | | | | | |
|---|---|---|---|---|---|
| 10 | 11 | 10 | 10 | 01 | 10 |
| 11 | 01 | 11 | 00 | 01 | 10 |
| 00 | 01 | 11 | 00 | 01 | 10 |
| 11 | 10 | 00 | 10 | 00 | 00 |
| 01 | 00 | 00 | 11 | 10 | 01 |
| 00 | 11 | 10 | 00 | 00 | 10 |

### Literacy
11, most illiterate

## Mortality
11, highest mortality

| 00 | 11 | 01 | 00 | 10 | 11 |
|----|----|----|----|----|----|
| 10 | 11 | 10 | 10 | 00 | 00 |
| 11 | 00 | 11 | 10 | 11 | 11 |
| 01 | 10 | 00 | 00 | 10 | 01 |
| 01 | 11 | 10 | 11 | 01 | 11 |
| 01 | 10 | 10 | 11 | 01 | 10 |

## Radio/ TV Penetration
11, least penetration

| 00 | 10 | 11 | 11 | 00 | 11 |
|----|----|----|----|----|----|
| 11 | 11 | 00 | 01 | 00 | 00 |
| 11 | 10 | 01 | 11 | 10 | 01 |
| 01 | 01 | 01 | 10 | 00 | 01 |
| 00 | 11 | 00 | 00 | 10 | 10 |
| 11 | 10 | 01 | 00 | 01 | 00 |

## Non-Structural Measures
11, no laws/ poor compliance

| 10 | 10 | 00 | 11 | 01 | 00 |
|----|----|----|----|----|----|
| 00 | 10 | 10 | 10 | 00 | 01 |
| 11 | 00 | 00 | 10 | 10 | 11 |
| 10 | 11 | 01 | 10 | 00 | 11 |
| 10 | 01 | 11 | 01 | 10 | 01 |
| 00 | 01 | 00 | 11 | 01 | 11 |

## Structural Measures
11, no structural measure

| 11 | 10 | 01 | 01 | 00 | 10 |
|----|----|----|----|----|----|
| 11 | 01 | 00 | 00 | 10 | 10 |
| 01 | 10 | 01 | 10 | 00 | 10 |
| 10 | 00 | 10 | 01 | 00 | 01 |
| 10 | 00 | 10 | 10 | 10 | 01 |
| 11 | 00 | 00 | 00 | 00 | 11 |

Using different set of weights used, it is seen that some of the results show similarities while some show differences. The result for urbanization, literacy rate and as well as structural measure are the same. The diagonal of the urbanization chromosomes show that half of the area are urbanized and half are not. Literacy rate and structural measure generated a model with more than half of the. As for the results on poverty it is shown that the result from using the previous weight is the reverse of what is generated when we use our own weight. Maybe because the of the different method we use and because they double the weight that they used for poverty. The remaining categories: mortality rate, radio/tv penetration, non-structural measures, also show that the first set of result is the inverse of what is generated in the second set of results. Further more, maybe it's because of the different approach in providing the weight that greatly affects the result that we gather from using the same program.

## SUGGESTIONS FOR FURTHER STUDY

The cost or 'penalty' function needs more research, and this might benefit from curve-fitting on available data. We could also explore bacteriologic algorithms, which work well with complex positioning problems.